\documentstyle[12pt,epsfig]{article}
\def\beqra{\begin{eqnarray}} 
\def\eeqra{\end{eqnarray}}
\def\beq{\begin{equation}}      
\def\eeq{\end{equation}}
\def\ds{\displaystyle}

\def\L{\Lambda}

\def\G{\Gamma}

\def \lta {\mathrel{\vcenter
     {\hbox{$<$}\nointerlineskip\hbox{$\sim$}}}}
\def \gta {\mathrel{\vcenter
     {\hbox{$>$}\nointerlineskip\hbox{$\sim$}}}}

{

\begin{document}
\sloppy
%
% title for draft version
%
\begin{titlepage}
\begin{flushright}
DFPD 97/TH/37\\
\end{flushright}
\vspace{3cm}
\centerline{\Large\bf Screening  Masses in SU(N) from}
\vspace{10pt}
\centerline{\Large\bf 
 Wilson Renormalization Group}
\vspace{24pt}
\centerline{\large Denis Comelli}
\begin{center}
{\it University of Lancaster, Lancaster LA1 4YB, UK}\\
{\it and}\\
{\it INFN, Sezione di Ferrara,\\
Via Paradiso 12, I-44100 Ferrara, Italy}
\end{center}
\vspace{10pt}
\centerline{\large Massimo Pietroni}
\begin{center}{\it Dipartimento di Fisica ``G.Galilei'',
\\University of Padova, Padova, Italy}
\end{center}
\begin{abstract}
We apply a gauge invariant formulation of Wilson Renormalization Group 
(RG) to the computation of the Debye and transverse gluon masses in pure
gauge $SU(N)$ at
high temperature. Following the Hard Thermal Loop effective field theory as a 
guideline, we develop an approximation scheme to the exact evolution 
equations. 

The Debye mass receives sizable corrections compared to the 
leading order perturbative result, mainly due to the infrared singular 
behavior in the transverse gluon sector. 
A non-vanishing mass for the transverse gluons
is found, which acts as an infrared 
regulator though not efficiently enough as 
to restore the validity of perturbation theory.
Indeed, discussing the role of higher dimensional operators, 
we show that the gauge coupling for the transverse modes typically flows to 
non-perturbative values unless extremely high temperatures are reached.

After comparing our results with recent lattice simulations, we
comment on the possibility of using this formulation of the RG as
a tool to construct an effective field theory for the non-perturbative, 
long wavelength, transverse modes.
\end{abstract}
\vspace{4.cm}
DFPD 97/TH/37\\
August 1997
\end{titlepage}
%\end{frontmatter}
%\def\baselinestretch{1.1}
{\bf \large 1.} The study of the properties of a 
non-abelian plasma at high temperature 
is of great interest both in cosmology, concerning the 
quark-hadron phase transition and the properties of the 
symmetric phase of the Standard Model, and in particle physics, in connection 
with the possibility of creating the quark-gluon plasma in heavy ion 
collisions. 

From a theoretical point of view, it has been known for a long time that 
perturbation theory is afflicted by
severe infrared problems caused by 
the static transverse gluons,  which remain massless at any order
\cite{linde}. As a consequence,
the loop expansion ceases to be an expansion
in powers of the coupling constant and, for instance, infinite loop orders 
contribute to the pressure at the same order $O(g^6)$. 

The Debye (or longitudinal) 
mass describes the screening of chromo-electrostatic fields in 
the plasma, and its numerical value influences the probability of $J/\Psi$ and 
$\Upsilon$ formation, among the main signatures of quark-gluon plasma
formation.  At leading order in perturbation theory it is given by
$m_L^2 = N g^2 T^2 / 3$ for $SU(N)$. 
At next to leading order infrared divergencies
show up, and the most one can compute is the coefficient of the 
term $g^{3}\log (m_L/m_T)$, $m_T$ being the transverse (or magnetic) 
gluon mass, which
at this level has to be introduced by hand \cite{rebhan}.  The coefficients of
the higher order terms in $g$ are not computable in perturbation theory.

In this letter we will apply the Wilson Renormalization Group (RG) method
introduced in  refs. \cite{noi1, noi2} to the computation of the Debye mass
$m_L$. As we have seen, the value of $m_L$ beyond leading order is influenced 
by the infrared cut-off in the transverse sector, $m_T$. 
We will first derive a coupled system of  RG flow equations for $m_L$ and $m_T$
describing the effect of the integration of thermal fluctuations at larger 
and larger length scales. Then, we will  include in the system also the 
running of the coupling constants and of wave function renormalizations.

The RG formulation developed in refs. \cite{noi1, noi2} is based on the 
introduction of
an infrared cut-off, $\L$,  in the thermal sector of the theory, so that the
modes of momentum $|\vec{k}| \gg \L$ are in thermal equilibrium at the
temperature $T$ while those 
with  $|\vec{k}| \ll \L$ are frozen at $T=0$.
The RG flow equations then interpolate between the full (renormalized)
quantum field theory at $T=0$ in the $\L \rightarrow \infty$ limit, and the 
full quantum field theory in thermal equilibrium at the temperature $T$ for 
$\L \rightarrow 0$. Compared to other formulations of the Wilson RG in the 
literature \cite{Pol}, we are then integrating out only the thermal 
fluctuations, all
the quantum fluctuations being already included in the initial
conditions of the RG flow. As discussed in detail in ref. \cite{noi2} the main 
advantage of introducing the cut-off only on the thermal sector is that BRST
invariance is preserved.
From a computational point of view, this allows us to use 
 Slavnov-Taylor (ST) identities 
as a powerful constraint in approximating the exact evolution equations.
From a physical point of view, we have a tool to derive an effective, 
gauge invariant, field theory, even for a non-zero value of the cut-off $\L$.

The effective field theory at the scale $g T$ has been developed 
in refs. \cite{HTL} and is known as Hard Thermal Loop (HTL). As external
momenta smaller than  $g T$ or next to leading order corrections
 are considered, 
the magnetic 
divergencies show up and the HTL resummation breaks down. 
Using the HTL effective theory as a starting point, the RG will guide us
deeper in the infrared. 
As we will see, as $\L$ approaches $m_T$ the gauge coupling in the
transverse sector rises to non-perturbative values, then large corrections
are to be expected to our results for $m_T$. On the other hand,
most of the renormalization of $m_L$ takes place for larger values of $\L$, 
where the couplings are still small, and the approximation to the RG equation 
is still reliable.

%The coarse grain differential equation o
%f the RG flow for the one loop effective potential is
%given by:
%\beqra
%\L\frac{\partial \:\:\:\:}{\partial \L} \Gamma_\L &=& -\frac{1}{2}\int 
%\frac{d^4 k}{(2 \pi)^4}\, \delta(|\vec{k}|-\L) \, N(|k_0|) \rho_\L(k) 
%\,\epsilon(k_0)
%\,\Gamma^{(3)}_\L(k, -k, 0)\,\\
%\Gamma_{\L=0}=\Gamma_{T=0}
%\label{flowtad}
%\eeqra
%
%Deriving with respect the  $\phi=A_{\mu}$  
%fields, we obtain an infinite system of differential equations
%describing the RG flow of the $\Gamma^{(n)}$ green function as a  function
%of the $\Gamma^{(n+2,n+1,n,...)}$  ones.
%\beqra
%\L\frac{\partial \:\:\:\:}{\partial \L} \Gamma_\L &=& -\frac{1}{2}\int 
%\frac{d^4 k}{(2 \pi)^4}\, \delta(|\vec{k}|-\L) \, N(|k_0|) \rho_\L(k) 
%\,\epsilon(k_0)
%\,\Gamma^{(3)}_\L(k, -k, 0)\,\\
%\Gamma_{\L=0}=\Gamma_{T=0}
%\eeqra
%where we recognize the Bose-Einstein distribution function, the
%full three-point function, and the spectral function $ 
%\rho_\L(k)$, defined as the discontinuity of the full propagator across 
%the real axis.
%The physical meaning of the flow equation (\ref{flowtad}) is evident: the new
%thermal modes coming into thermal equilibrium at the scale $\L$ are 
%weighted by the full spectral function, induced
%by all the quantum and thermal modes already integrated out.
%\section{Gauge invariance}

\vspace{0.5 cm}{\bf \large 2.}
 In this letter we will be mainly interested in the longitudinal 
(or Debye) and 
transverse (or magnetic) masses, defined as the static poles of the full 
propagator, 
\beq
m_{L,T}^2 = \Pi_{L,T}(q_0=0, |\vec{q}|^2=-m_{L,T}^2)\,,
%\nonumber\\
%m_T^2 &=&\Pi_T(q_0=0, |\vec{q}|^2=-m_T^2)\,,
\label{masse}
%\eeqra
\eeq
where $\Pi_{L,T}$ are obtained from the self-energy $\Pi^{\mu \nu}$ as
$\Pi_L=\Pi^{00}$ and $\Pi_T=-1/2\, \Pi^{ii}$.
As shown in  ref.~\cite{KKR}, the definitions (\ref{masse}) are
gauge-independent. The same holds even in presence of the infrared cut-off,
since it  does not break BRST invariance.

The RG flow of the Debye and magnetic masses is described by the
equations \cite{noi2}:
\beq
 \L\frac{\partial \:\:\:\:}{\partial \L} m_{L,T}^2 = 
\left. \frac{\L\frac{\partial \:\:\:\:}{\partial \L} \Pi_{L,T}(q)}
%B^{\mu \nu}(q)
%F_{\mu \nu}(q)}
{1+\frac{\partial \Pi_{L,T}(q)}{\partial|\vec{q}|^2}}
\right|_{|\vec{q}|^2=-m^2_{L,T}}  \,,
\label{flows}
\eeq
%\ds \L\frac{\partial \:\:\:\:}{\partial \L} m_{T}^2 &=&
%\ds \left. \frac{\L\frac{\partial \:\:\:\:}{\partial \L} \Pi_T(q)}
%\frac{1}{2} A^{\mu \nu}(q)F_{\mu \nu}(q)}
%{1+\frac{\partial \Pi_T(q) }{\partial|\vec{q}|^2 }}
%\right|_{|\vec{q}|^2=-m^2_T}
%\label{flows}
%\eeqra
where the flow equations for $\Pi_{L,T}$ are obtained by taking the 
appropriate 
components of that for the self-energy. 

In the case of one particle irreducible vertices the RG flow equations 
can be obtained by the 
following simple recipe:
1) take the expression for the one loop correction to the desired vertex;
2) substitute the tree level propagators and vertices in it with the full, 
cut-off dependent ones;
3) take the derivative with respect to the explicit cut-off dependence 
({\it i.e.} derive only the cut-off function in the propagators).
For the self-energy we get \cite{noi2}
\beqra
\ds \L\frac{\partial \:\:\:\:}{\partial \L}\Pi_{\L, \mu \nu}(q)=
\ds -\frac{i}{2} \int \frac{d^4k}{(2 \pi)^4}K_{\L, \rho \lambda}(k)\;
[\Gamma_{\L, \mu\nu\rho \lambda}(q,-q,k,-k) \quad\quad\quad\quad\quad
\quad&&\nonumber\\
\ds +2 \, 
\Gamma_{\L, \mu\rho\eta}(q,-k,k-q)\; G_{\L, \eta \delta}(k-q)\;
\Gamma_{\L, \nu \lambda \delta }(-q,k,q-k)\;] + {\rm Ghost}.&&
\label{run}
\eeqra
The kernel,
\[
K_{\L\,,\rho \lambda}(k) = \rho_{\Lambda, \rho\lambda}(k) 
\varepsilon(k_0) \L \,\delta(|\vec{k}|-\L)
\, N_b(|k_0|)\,,
\]
contains the Bose-Einstein distribution function, $N_b(|k_0|)$,
and the full spectral function 
\[ \rho_{\Lambda, \rho\lambda}(k)= 
-i\varepsilon(k_0) {\rm Disc} G_{\L,  \rho\lambda}(k)\;,
\]
and $G_{\L,  \rho\lambda}$, $\Gamma_{\L, \mu\rho\eta}$, and 
$\Gamma_{\L, \mu\nu\rho \lambda}$, are the full propagator and vertices computed
at the scale $\L$.

Up to now no approximation has been performed, and the evolution equations 
(\ref{flows}) define the screening masses non-perturbatively. Of course, in 
order to solve them, one should know the exact propagator and vertices 
(including their momentum dependence) for any value of the cut-off 
$\L$. This requires managing an infinite system of differential equations 
for the self-energy and all higher order vertices, which is equivalent 
to solving finite temperature QCD. 

As discussed for instance in \cite{Pol}, RG equations can be solved 
iteratively, reproducing the usual loop expansion.
Introducing in the RHS the $\L$-{\it independent} 
expressions for the propagator  and vertices at the loop {\it l} and 
integrating in $\L$ yields the results at the {\it l+1} loop order.
In this case, we recover the well known results $m_L\equiv m_L^{LO}= \sqrt{N/3}
g T$ 
and $m_T=0$
after the first iteration. 
However it is well known that in high temperature QCD the loop expansion 
ceases to be a sensible 
approximation already at the second loop, since contributions $O(g^{3/2} T)$ 
to $m_L$ are generated at each higher order \cite{HTL,rebhan}. 
Concerning $m_T$,
it is easy to realize that starting from $m_T=0$, no non-vanishing value can
be generated at any order in the loop expansion.

Analogously, the HTL loop expansion at {\it l+1} order 
can be reproduced by introducing the  $\L$-{\it independent} results obtained
from the HTL effective theory at the {\it l}-loop in the RHS of 
eqs. (\ref{flows}). After the first iteration, the result of eq. (\ref{nlo}) 
for $m_L$ discussed 
by Rebhan in ref. \cite{rebhan} is obtained.

In order to go beyond the HTL effective field theory and closer to the RG
framework  the next logical step is 
to promote the  $\L$-{\it independent} HTL propagator and vertices to 
 $\L$-{\it dependent} ones. Then, the contribution to $m_{L,T}$ at the
scale $\L$ will be given by the RHS of eqs. (\ref{flows}) with 
$m_{L,T}$ (and the coupling constant) evaluated at the same
scale $\L$. Moreover, we will require that the improved vertices and propagator
satisfy the same tree level ST identities as the HTL ones \cite{HTL}.

The HTL propagator and vertices are complicated  functions of the momenta 
\cite{HTL}, whose expressions simplify when the  external energy vanishes.
Since in this letter we are interested in computing the static 
quantities in (\ref{masse}) a great simplification to the 
flow equations can be obtained by rotating to the imaginary time 
\cite{landsman}. The 
evolution equation (\ref{run}) then takes the form (from now on we omit the 
$\L$-dependence of the various quantities)
\beqra
 \L\frac{\partial \:\:\:\:}{\partial \L}\Pi_{\mu \nu}(q_0=0,|\vec{q}|)&=&
T \sum_n F_{\mu\nu}(z=2 i \pi n T;|\vec{q}|; m_L, m_T) \nonumber\\
&-& 
\int\frac{dk_o}{2\pi}  F_{\mu\nu}(k_0;|\vec{q}|; m_L, m_T)
\label{flowit}
\eeqra
with
\beqra
 F_{\mu\nu}(k_0;|\vec{q}|; m_L, m_T) = 
-\frac{1}{2}\int\frac{d^3 k}{(2 \pi)^3} \L \,\delta(|\vec{k}|-\L) 
G_{\rho\lambda}(k) \left[
\Gamma_{\mu\nu\rho \lambda}(q,-q,k,-k) \quad\quad\quad \right.&&\nonumber\\
\left.\ds +2 \, 
\Gamma_{\mu\rho\eta}(q,-k,k-q)\; G_{\eta \delta}(k-q)\;
\Gamma_{\nu \lambda \delta }(-q,k,q-k)\;\right] + {\rm Ghost}\;.&&
\nonumber
\eeqra
Notice that, due to the cancellation between the Matsubara term in the first
line and the ``$T=0$'' term in the second line,  eq. (\ref{flowit}) is free from
ultraviolet divergencies, as it is manifest also in the form (\ref{run}), 
which  contains the Bose-Einstein function.
This is of course to be  expected, since these RG equations describe the 
effect of thermal fluctuations only.

The evolution equation for $m_L$ involves $F_{00}$. We approximate it by using 
{\it tree level} propagator and vertices for the non-zero Matsubara 
modes and in the  ``$T=0$'' part,
\[
F_{00}(k_0;|\vec{q}|; m_L, m_T) \simeq F_{00}(k_0;|\vec{q}|=0; m_L=0, m_T=0)
 \quad\quad{\rm for} \:\:n\neq 0\;.
\]
Analogously, in the evolution equation for $m_T$, 
we approximate $F_{ii}$ as
\[
F_{ii}(k_0;|\vec{q}|; m_L, m_T) \simeq F_{ii}(k_0;|\vec{q}|=0; m_L,  m_T=0)
 \quad\quad{\rm for} \:\:n\neq 0\;.
\]
These 
approximations will lead to $O(g^2 m_L^2/T^2)$ and $O(g^2 m_T^2/T^2)$ 
errors, respectively, since $O(g^2 m_{L,T}/T)$ terms may only come from the, 
infrared problematic, zero
Matsubara mode.

The next-to-leading correction in the  HTL resummed perturbation theory can
be obtained by using resummed propagators and vertices for the zero mode only
\cite{rebhan}.
In the same spirit, we will employ HTL-inspired propagator and vertices only 
for $n=0$.

The propagator we need is
\beqra
\left.\Delta_{\mu\nu}\right|_{k_0=0} &=&
%\Delta_{\mu\nu}^{L}+\Delta_{\mu\nu}^{T}+\Delta_{\mu\nu}^{gauge}=\nonumber\\
\frac{Z_L}{|\vec{k}|^2+m_{L}^2} 
\,g_{\mu 0}\, g_{\nu 0} \nonumber\\
&&+ \frac{Z_T}{|\vec{k}|^2+m_{T}^2} \left(g_{\mu\nu} - 
g_{\mu 0}\, g_{\nu 0} + \frac{k_\mu\, k_\nu}{|\vec{k}|^2}\right) 
\left.-\alpha \frac{k_\mu \, k_\nu}{|\vec{k}|^4}\right|_{k_0=0}\,,
\label{prop}
\eeqra

where, compared to the true HTL propagator \cite{rebhan} 
we have $\L$-dependent 
$m_{L,T}$ and $Z_{L,T}$.
The wave function renormalizations are defined, as usual, as 
$Z_{L,T}= 1 + \left.\frac{\partial\:\;}{\partial |\vec{q}|^2}\Pi_{L,T}
(q_0=0,|\vec{q}|^2) \right|_{|\vec{q}|^2
=m_{L,T}^2}$.

In the HTL approximation, the three and four gluon vertices, evaluated
at zero external energy, reduce to their tree level expressions. Here, the 
presence of a non-zero $m_T$ forces us to modify the trilinear vertex in order
to preserve the gauge invariance of the evolution equations (\ref{flows}).
Requiring that the new vertex is related to the propagator (\ref{prop}) by the
tree-level ST identity does not fix it completely. The transverse component of
the vertex could be in principle determined by studying its RG equation. In  
order to limit the number of flow equations, in this letter we 
will instead fix the transverse part by imposing that the improved vertex has
the same form as that obtained by Alexanian and Nair \cite{Nair},
\beq
\left.\Gamma_{\mu\nu\rho}(q,k,-q-k)\right|_{q_0=k_0=0}=
g [v_{\mu\nu\rho}(q,k,-q-k)+
V_{\mu\nu\rho}(q,k,-q-k)]
\label{trilinear}
\eeq
where
\[
v_{\mu\nu\rho}(q,k,-q-k)=
(q-k)_{\rho} g_{\mu\nu} +(2k+q)_{\mu} g_{\nu\rho} -(2q+k)_{\nu} g_{\mu\rho} 
\]
and
\beqra
V_{0\nu\rho}(q,k,-q-k)&=&
V_{\mu0\rho}(q,k,-q-k)=
V_{\mu\nu0}(q,k,-q-k)=0 \,,\nonumber\\
V_{ije}(q,k,-q-k)&=&-m_T^2\, A\,[B\, k_i\, k_j\, k_e\,+\,
C\,(q_i\,q_j\,k_e
+\,q_j\,q_e\,k_i\,+\,q_i\,q_e\,k_j)\nonumber\\
&&-\,(q\leftrightarrow k)]
\nonumber
\eeqra
with
\[
A=\frac{1}{k^2\,q^2-(q\,k)^2}\,,\;\;\;\;
B=\frac{q\,k}{k^2}-\frac{q(q+k)}{(q+k)^2}
\,,\;\;\;\;C=\frac{k(q+k)}{(q+k)^2}\;.
\]
Note that the vertex correction $V_{\mu\nu\rho}$ does not appear
in the evolution equation for $m_L$ but only in that for $m_T$.
Inserting eqs. (\ref{prop}), (\ref{trilinear}),
and the tree level expressions for the four-gluon vertex and the ghost
propagator and vertices in the on-shell evolution equations (\ref{flows})
the gauge parameter dependence drops, as can be checked by explicit
calculation.

In the flow equations (\ref{flows}) the vertices are evaluated at 
external momenta of order $\L$. As a first step, the thermal 
contribution to the renormalization of the coupling constant will be neglected,
and the  dependence on the external momentum will be taken into account by
using the $T=0$ beta function at two loops,
\beq
g^{-2}(\L) = \frac{11\,N}{24 \pi^2} \log\left(\frac{\L}{\L_{\overline{MS}}}
\right) +
        \frac{17\,N}{88 \pi^2}
\log\left[2 \log\left(\frac{\L}{\L_{\overline{MS}}}\right)\right]\;
\label{coupling}
\eeq
for $\L > m_L$, and $g(\L)=g(m_L)$ for $\L \le m_L$.
The wave function renormalizations $Z_{L,T}$ will be fixed at $Z_{L,T}=1$. 

The thermal renormalization of $m_L/T$ as a function of the cut-off $\L$ 
can be read from Fig.1.
\begin{figure}
\begin{center}
\mbox{\epsfig{file=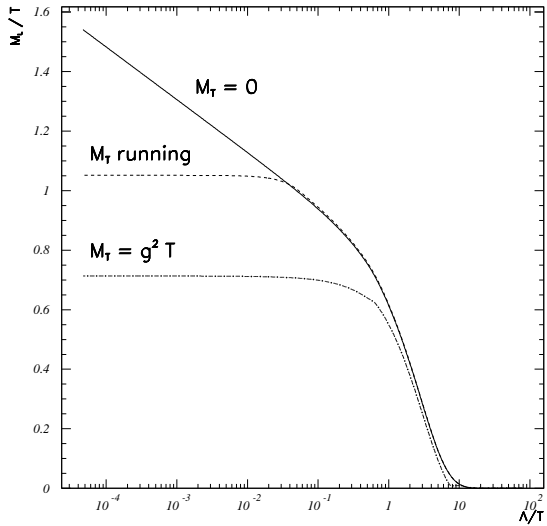,width=8.truecm}}
\end{center}
%\hbox{\epsfig{file=glgt08log.ps,width=2.1in}}}
\caption{Running of $m_L/T$ vs. $\L/T$ in $SU(3)$ for $T = 10^4
{\L_{\overline{MS}}}$. The solid line is obtanied fixing  $m_T=0$, and the 
dash-dotted line for  $m_T= g^2(T) T$. The dashed line corresponds to the 
solution of the system of RG equations for $m_L$ and $m_T$.}
\end{figure}
The effect of the fluctuations integrated out at 
scales $\L/T \gta O(10)$  is negligible due to Boltzmann suppression. 
Around $\L/T = O(1)$ the HTL contribution
 is rapidly built up. Stopping the running at this scale 
basically reproduces the HTL effective field theory results. Going to 
lower energy scales the infrared divergencies in the magnetic sector 
start showing up and, depending on the value of $m_T$, their effect  
may  eventually  overcome the leading perturbative result. This is 
clearly seen from the  upper  line, where  we have set $m_T=0$.

In next-to-leading order perturbative computations the generation of a 
$O(g^2 T)$ magnetic mass by non-perturbative effects is usually invoked as a 
mechanism to regulate the infrared divergencies coming from the
transverse gluon sector. The computation of $m_L$ at next-to-leading 
order gives \cite{rebhan}
\beq
m_L^2 = \frac{N}{3} g^2 T^2 \left( 1 +\frac{N}{2\pi} g \log\frac{m_L}{m_T} 
+ C g + O(g^2)
\right)
\label{nlo}
\eeq
where $C$ is a coefficient receiving contributions at any loop order and 
therefore not computable in perturbation theory. If we introduce by hand an 
{\it ad hoc} $m_T=
O(g^2 T)$, then perturbation theory can tell us the coefficient of the 
$g \log(1/g)$ term, but in any case it becomes meaningless beyond this order.

Fixing $m_T=g^2(T) T$, where $g(T)$ is obtained by computing 
eq. (\ref{coupling}) at the thermal scale of perturbation theory, 
$\L_{th}= 4 \pi T \exp(-\gamma_E) \simeq 7.055\, T$ \cite{KajantieDR},
the lowest line in Fig.1 is obtained.
In Fig. 2  $m_L/T$ is  plotted as a function of the temperature.
The lowest line corresponds again to $m_T=g^2(T) T$. Notice that it agrees with
the next-to-leading order perturbative result (eq. (\ref{nlo}) with $C=0$) 
only for very high temperatures $T \gta 10^{10} \L_{\overline{MS}}$.
%and compare it with
%eq. (\ref{nlo}), which best fits the data for $C=1.66$. Notice how the 
%agreement is acceptable only up to $g \lta 0.4$, or $T/\L_{\overline{MS}} 
%\gta 10^{10}$.
\begin{figure}
\begin{center}
\mbox{\epsfig{file=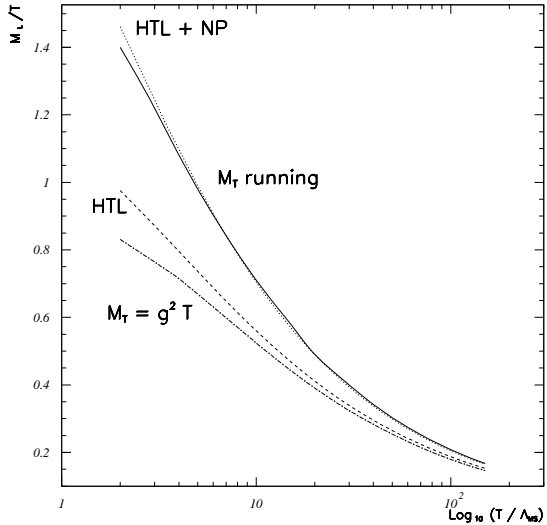,width=8truecm}}
\end{center}
%\hbox{\epsfig{file=glgt08log.ps,width=2.1in}}}
\caption{Plot of  $m_L/T$ vs. $\log_{10} (T/\L_{\overline{MS}})$ 
for $SU(3)$. The lowest 
line is obtained by fixing $m_T=g^2(T) T$, the ``HTL'' line represents 
eq.(\ref{nlo}) computed for $g=g(T)$ and $C=0$, 
and the ``HTL + NP'' (dotted line) is the same equation  but for 
$C = 1.3$. The solid line is the result of the RG system for $m_L$ and $m_T$.}
\end{figure}
The next step is to let $m_T$ run, studying the coupled system of 
RG equations for $m_L$ and $m_T$. 
We note immediately that, at this level of the approximation,
$m_T=0$ is a fixed point  of the RG flow. In fact, in this case, the RHS of the
flow equation reduces to the one-loop expression which, upon integration, 
gives $m_T=0$.

The possibility of  a thermal renormalization of the magnetic mass 
is then related to its initial value at $T=0$. The 
structure of the gluon propagator is still  subject of intense study, 
and no definite answers are available. However different groups have found 
evidence for the existence of a pole at non-zero momentum of 
$O(\L_{\overline{MS}})$  
\cite{poleglue}. In the following we will  assume that 
such a pole indeed 
exists, and will use $m_T = m_L = \eta \L_{\overline{MS}}$, 
with $\eta$ a coefficient 
of $O(1)$, as initial value for the running. Increasing $T$, the 
sensitivity to $\eta$ becomes weaker and weaker, since 
changing this parameter is equivalent to rescaling $\L_{\overline{MS}}$, which 
enters the problem only in the argument of the logarithm in eq. 
(\ref{coupling}). 

The result is plotted with the continuos line in Fig. 2. A comparison with
the next-to-leading expression of eq. (\ref{nlo}) (dotted line), computed for
$g=g(T)$,
allows us to determine the non-perturbative coefficient $C$ as
\[C = \left\{ \begin{array}{c}
1.1 \;\;\;\;\;\;{\rm for\;\; SU(2)}\\
1.3 \;\;\;\;\;\;{\rm for\;\; SU(3)}\\
\end{array}
\right.\;.
\]
As a general fact, we observe that 
the corrections to $m_L$ are larger than those obtained by fixing 
$m_T=g^2 T$. This is due to the fact that the values of $m_T$ given by 
the RG are generally much smaller than  $g^2 T$, so the decoupling 
of the RG flows takes place deeper in the infrared. It can be seen
explicitely from  the dashed line in Fig. 1.
A $m_T=O(g^2T)$ naturally emerges from gap equations, 
\cite{buch,Nair} where 
the $\L$-dependence of $m_T$ inside the loop
integral is neglected.
Once this dependence is kept, the simple scaling law $m_T=O(g^2T)$
is lost. We can understand it by looking at the approximate form of the
flow equation for $m_T$ obtained in the limit $\L\gg m_T$,
\beq
\L \frac{\partial\quad}{\partial \L} m_T^2 = - K g^2 T \frac{m_T^2}{\L}\;,
\label{appr}
\eeq
with $K=55/8 \pi^2 \simeq 0.697$ in the case of $SU(3)$.
We may also neglect the $\L$-dependence of $g$ and the 
contribution of the running for $\L$ below $m_T$, as 
we can verify by comparison with the full numerical results.

The gap equation approximation corresponds to integrating eq. (\ref{appr})
with a $\L$-independent $m_T=m_{T, gap}$ such that
\beqra
\frac{m_{T, gap}^2}{T^2} &=& \frac{m_{T}^2(\L=\infty)}{T^2} + 
K g^2 \frac{m_{T, gap}^2}{T} \int_{m_{T, gap}}^\infty \frac{d\L}{\L^2}
\nonumber\\
&\simeq & K g^2 \frac{m_{T, gap}}{T}\;\quad\quad\quad(T\gg 
m_{T}(\L=\infty)= \eta \L_{\overline{MS}})\;,
\nonumber
\eeqra
leading to the non-vanishing solution $m_{T, gap}= K g^2 T$.

On the other hand, integrating eq. (\ref{appr}) down to 
$\L=m_T$ 
without further approximations, gives the equation
\beq
m_{T}^2 = m_{T}^2(\L=\infty) \exp\left(K g^2 T/m_{T}\right)\;,
\label{RGgap}
\eeq
which always gives $m_{T} \ll m_{T, gap}$ for $T\gg \eta\,\L_{\overline{MS}}$.

From the full RG equations we obtain values for $m_T$ which are in reasonably 
good  agreement with  the solutions of eq. (\ref{RGgap}) and, in the 
range of 
temperatures considered, are very well approximated by the law (in $SU(3)$)
\[
m_T = 0.128 \,g^2 \,T (1 + 2.26 \,g \log g)\; .
\]
\\

\vspace{0.5 cm}
{\bf \large 3.} In order to assess the reliability of our approximation 
to the RG flow 
equations for $m_L$ and $m_T$ we must consider the thermal running of higher 
dimensional operators. In particular, we will study the running of the 
gauge coupling constant in order to determine the efficiency  of the 
screening masses as infrared regulators.

In the flow equations for $m_L$ and $m_T$ only two types of trilinear 
vertices appear, $\G_{00i}$ and $\G_{ijk}$, where $i,j,k = 1,2,3$. 
The RG equations for these two vertices  differ remarkably in the 
infrared, since that for $\G_{ijk}$ receives contribution from the loop 
in which the three circulating  gluons are all transverse, of mass $m_T$, 
whereas in the loops contributing to the running of $\G_{00i}$ at least 
one longitudinal gluon appears. As a consequence, it is convenient to modify 
the parameterization of the trilinear vertex in eq. (\ref{trilinear}) by 
introducing  two different  coupling constants, $g_L$ 
for $\G_{00i}$, and $g_T$ for $\G_{ijk}$. Of course, due to the presence of the 
thermal bath, this is not in contradiction with Lorentz invariance.

From the RG flow of $\G_{\mu\nu\rho}$ we extract those of $g_{L,T}$, 
defined as
\beqra
g_L&=&\left.\lim_{q\rightarrow k}\frac{(q+k)^{\rho}}{q^2-k^2}
 \Gamma_{0\,0\,\rho}[q,k,-q-k]\right|_{q_0=0 \,, |\vec{q}|^2=
-m_L^2}\,,\nonumber\\
g_{T}&=&\left. \lim_{q\rightarrow k}\frac{(q+k)^{\rho}}{2(q^2-k^2)}
 \Gamma_{i\,i\, \rho}[q,k,-q-k]\right|_{q_0=0 \,, 
|\vec{q}|^2=-m_T^2}\,,
\label{acco}
\eeqra

The running of the couplings is shown in Fig. 3.
\begin{figure}
\begin{center}
\mbox{\epsfig{file=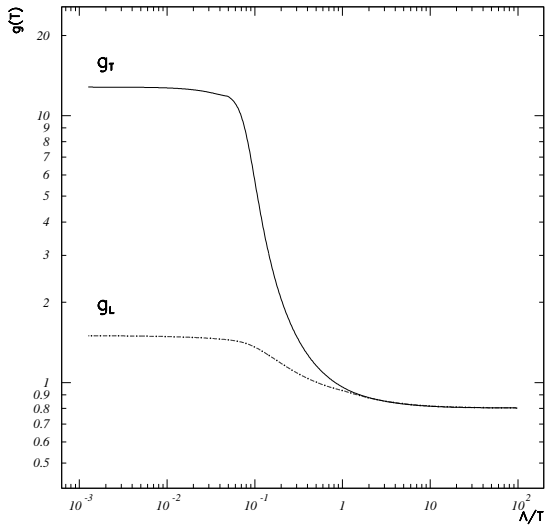,width=8.truecm}}
\end{center}
%\hbox{\epsfig{file=glgt08log.ps,width=2.1in}}}
\caption{RG flows for  $g_L$ e $g_T$ defined as in eq. (\ref{acco}) for $SU(3)$
and $T=10^4 \L_{\overline{MS}}$.}
\end{figure}
Notice the impressive difference, induced by thermal effects, between  
$g_L$ and 
$g_T$.
 This is due to the fact that graphs with only transverse gluons 
in the loop contribute to the running of $g_T$ but not the ones 
of $g_L$. When $m_T\ll \L \ll T$ the transverse sector of the theory is in 
the three-dimensional  regime, 
and the coupling constant $g_T$ grows as $1/\L$ in the infrared.
The non-vanishing $m_T$ leads to  decoupling for $\L \lta m_T$, 
but it takes place when the coupling $g_T$  has already 
grown to 
non-perturbative values. 

Comparing Figs. 1 and 3 we see that the evolution of the coupling constants 
affects our results for $m_L$ and $m_T$ in a different way. First, notice
that in the flow equation for $m_L$ only the vertex $\G_{00i}$, and then the
safer coupling $g_L$, appears explicitly. 
The effect of $g_T$ is felt by $m_L$ only indirectly, through
the running of $m_T$. Moreover, the running of $m_L$ is almost completely 
decoupled
before $g_T$ starts its crazy rise. As a consequence, the non-perturbative 
regime for $g_T$ does not affect very much our results for $m_L$, since 
there is no time enough to communicate it to the longitudinal sector.

On the other hand, since most of the running of $m_T$ takes place when $g_T$ 
is also running the inclusion of higher order operators
gives larger corrections to $m_T/T$. We estimate 
that they are typically of $O(1)$,
 corresponding to a few percent corrections to $m_L/T$.\\

\vspace{0.8 cm}
{\bf \large 4.} Summarizing, we have employed the Wilson Renormalization Group
at finite temperature introduced in refs. \cite{noi1,noi2} to study  the 
gluon self-energy in high temperature QCD. In order to keep contact with
the perturbative and HTL-resummed results, we have developed an approximation
scheme based on the use of HTL-inspired Ans\"{a}tze for the propagator
and vertices appearing in the RG equations.

We have concentrated on $m_L$ and $m_T$, the Debye and magnetic masses, 
deriving a system of differential equations for them, the coupling constants 
and the wave function renormalizations. 
We obtain large corrections to $m_L$ with respect to the leading order result,
$m_L^{LO}=\sqrt{N/3} \,g T$, mainly due to infrared effects 
in the transverse sector.
 
Our results for $m_L$ in pure gauge $SU(2)$ are in good agreement with the 
lattice results of ref. \cite{heller} on the pole mass of the longitudinal
gluon propagator in the Landau gauge. For instance, for $T/\L_{\overline{MS}}=
4000$ we find $m_L/T= 1.12$ whereas they obtain $m_L/T=1.1 - 1.3$, the leading
order perturbative value being $m_L/T \simeq 0.84$.  
Lattice computations of the screening masses have been  reported  also by 
Kajantie et al. in ref. \cite{Shap}, where  the long
distance behaviour of certain gauge-invariant operators is studied. In this 
case corrections  a factor two to three larger than ours and those of
ref. \cite{heller} are generally found. 
As discussed in \cite{karsch,heller} a possible explanation of this 
discrepancy  could be that  
the effective thermal mass extracted from  gauge-invariant composite 
operators arises from the superposition of several decoupled gluons.

Concerning $m_T$,  results much smaller than 
the value  $g^2 T$ 
obtained from gap equations are typically found, and this has been ascribed to
the scale dependence of $m_T$.

The inclusion of the running of the coupling constants typically leads to 
sizable corrections in the transverse sector, since the relevant coupling rises
to non-perturbative values in the infrared.
This result clearly indicates  that the  dynamics of the  transverse 
modes of wavelength $\gta 1/m_T$ is highly non-perturbative. 

In perturbation
theory, the non-perturbative  modes circulate in the loop at any order, 
leading to the infrared problem that we have already discussed. The same
statement that the HTL give the effective theory at the scale $g T$ is 
meaningful only at leading order, since at higher orders the longer scales come
into play.
On the other hand, in a RG framework, the scale $\L$ can be used to keep 
the long wavelength modes under control, allowing the definition of  
effective theories at larger and larger scales in a clean way. 
This procedure usually runs into problems for gauge theories, since the 
introduction of the cut-off $\L$ breaks gauge invariance. In our framework 
this is not the case, since the cut-off is imposed on the thermal, on-shell,
sector of the theory only \cite{noi2}. 
The construction of effective field theories along 
the lines illustrated in ref. \cite{muller} for the scalar theory can then be
properly carried out in this context.

\vspace{0.4in}
\centerline{\bf Acknowledgments}
We are very grateful to our friend and collaborator M. D'Attanasio for 
participating to the early stages of this work and for many 
inspiring discussions afterwards. 
D.C. acknowledges funding from a PPARC research grant, and part of this 
work was done during a visit of M.P. to Lancaster which was partially funded 
by a PPARC visiting fellowship grant.

%\begin{figure}
%\begin{center}
%\end{center}
%\caption{.}
%\end{figure}

\end{document}